\documentclass[aps,prd,preprintnumbers]{revtex4}
\usepackage{amsfonts,amsmath,units,amssymb,graphicx,bm,verbatim,longtable,mathrsfs,subfigure,mathtools,braket,slashed}
\usepackage[colorlinks=false,urlcolor=blue]{hyperref}
\allowdisplaybreaks

\newcommand{\eV}{\textrm{eV}}

\newcommand{\GeV}{\textrm{GeV}}
\newcommand{\TeV}{\textrm{TeV}}
\newcommand{\BM}{\left(\begin{array}}
\newcommand{\EM}{\end{array}\right)}
\begin{document}
\title{Neutrinos with a linear seesaw mechanism in a scenario of gauged $B-L$
    symmetry}
\author{Claudio O. Dib${}^{}$}
\email{claudio.dib@usm.cl}
\author{Gast\'on R. Moreno${}^{}$}
\email{gaston.moreno@postgrado.usm.cl}
\author{Nicol\'as A. Neill${}^{}$}
\email{nicolas.neill@gmail.com}
\affiliation{${}^{}$Universidad T\'ecnica Federico Santa Mar\'{\i}a\\
and\\
Centro Cient\'{\i}fico-Tecnol\'ogico de Valpara\'{\i}so\\
Casilla 110-V, Valpara\'{\i}so, Chile}
\begin{abstract}
We consider a mechanism for neutrino mass generation, based on a local $B-L$ extension of the standard model, which becomes a linear seesaw regime for light neutrinos after spontaneous symmetry breaking. The spectrum of extra particles includes heavy neutrinos with masses near the TeV scale and a heavy $Z'$ boson, as well as three extra neutral scalars and a charged scalar pair. We study the production and decays of these heavy particles at the LHC. $Z'$ will decay mainly into heavy neutrino pairs or charged lepton pairs, similar to other low scale seesaw scenarios with local $B-L$, while the phenomenology of the extra scalars is what distinguishes the linear seesaw from the previous models. 
One of the neutral scalars is produced by $Z' Z'$ fusion and decays mainly into vector boson pairs, the other two neutral scalars are less visible as they decay only into heavy or light neutrino pairs, and finally  the charged scalars will decay mainly into charged leptons and missing energy.
\end{abstract}
\maketitle

\section{Introduction}\label{sec:intro}

The experimental evidence that neutrinos have non vanishing masses and their mass eigenstates are admixtures of the lepton flavors is an indication of new physics beyond the Standard Model (SM) \cite{GonzalezGarcia:2002dz}. In the SM, neutrinos are massless left handed particles, while their right handed components, $\nu_R$, are absent. As neutrinos are massive, most if not all possible explanations lead us beyond the SM. For example, if we insist in having SM fields only, the neutrino mass must be of Majorana type, $ m_\nu\, {\nu_L} \nu_L$. However, since the neutrino fields are part of and $SU(2)_L$ doublet, these Majorana mass terms can only arise from effective non-renormalizable terms of the form $\sim (L\phi)(L\phi)/M$, where $L$ and $\phi$ are the lepton and Higgs doublet respectively \cite{Weinberg:1979sa}. We are thus led to introduce an effective scale $M$ which indicates extra physics. Alternatively,  if we include the right handed singlets $\nu_R$ to construct Dirac mass terms from standard Yukawa interactions, we cannot avoid including a Majorana term of the form $M\nu_R \nu_R$ as well, again introducing a new scale $M$ and the existence of heavy Majorana states.
We can forbid this Majorana term if we impose Lepton Number conservation; however, since Lepton Number in the SM is just a global symmetry, it can be considered to appear accidentally in the formulation instead of being a fundamental and necessary symmetry.     

An elegant extension of the SM that includes neutrino masses is based on the assumption that $B-L$ is actually a local --or gauge-- $U(1)$ symmetry, spontaneously broken due to an extended Higgs sector \cite{Mohapatra:1980qe,Wetterich:1981bx,Buchmuller:1991ce,Khalil:2006yi}. 
The motivation for promoting  $B-L$ to a local symmetry is manifold: gauge theories are successful quantum theories; local symmetries are a fundamental piece of the dynamics; finally, in local $B-L$, $\nu_R$ is not just introduced at will,  but is required by the anomaly cancellation of the $B-L$ current. In its minimal version \cite{Khalil:2006yi, Basso:2008iv}, a local $B-L$ model contains one extra neutral vector boson, $Z^\prime$, and one extra singlet scalar. The mass matrix of the neutral leptons in this case, in the basis $(\nu_L^{\ c}, {\nu_R})$,  resembles that of a seesaw model of type I 
\cite{Minkowski:1977sc, Yanagida:1979as, GellMann:1980vs, Mohapatra:1979ia},
\begin{equation}
{\cal L}_m =  - \frac{1}{2}  (\overline{\nu_L}, \overline{\nu_R^{\ c}})  \left( \begin{array}{cc}
0 & m_D \\
m_D^T & M_R \end{array} \right)  \left( \begin{array}{c}
\nu_L^{\ c} \\
\nu_R
 \end{array} \right) + h.c.
\label{type1}
\end{equation}
In this original seesaw model for neutrino masses, 
the light neutrinos get lighter masses, $m_\nu \sim m_D^2/M_R$ just as the heavy modes get heavier, namely $\sim M_R$. At the same time, the heavy-to-light mixings get smaller: $\theta \sim m_D/M_R$. If one assumes that $m_D$ is comparable to the electron mass ($\sim 1$  MeV), the light neutrino masses in this model will be of the size indicated by oscillation experiments ($\sim 10^{-2}$ eV) provided $M_R\sim 100$ TeV, and therefore $\theta \sim 10^{-8}$. Consequently, heavy neutrinos from this model have very small mixing with the SM sector and have masses well above the reach of the LHC. Moreover, if we assume $m_D$ to be near the electroweak scale, $10^2$ GeV, which corresponds to Yukawa couplings $\sim 1$, in that case $M_R\sim 10^{15}$ GeV, close to the GUT scale and far out of reach from any foreseen experiment. 

In other types of scenarios, called \emph{low scale seesaw}, the smallness of the light neutrino masses is not just due to the presence of very heavy states, but to another mass parameter which in some sense is naturally small. These scenarios are inspired on grand unified theories and superstring models, which include two extra neutral lepton singlets per family \cite{Wyler:1982dd, 
Witten:1985xc, Mohapatra:1986bd},  $\nu_R$ and $S$. In the basis $(\nu_L^{\ c}, \nu_R, S)$,  the original mass matrix of these models is of the form
\begin{equation}
{\cal L}_m =  - \frac{1}{2}  (\overline{\nu_L}, \overline{\nu_R^{\ c}}, \overline{S^c})    \left( \begin{array}{ccc}
0 & m_D &0\\
m_D^T &0 & M_R\\
0 & M_R^T & 0 \end{array} \right)  
\left( \begin{array}{c}
\nu_L^{\ c} \\
\nu_R
\\
S
 \end{array} \right) + h.c. 
\label{lowscale}
\end{equation}
With this mass texture, the light neutrinos are actually massless and lepton number is conserved. However, one can include a naturally small mass parameter that violates lepton number and generates  small masses to the otherwise massless neutrinos, without requiring $M_R$ to be extremely large (hence the name \emph{low scale} seesaw). 

If this small parameter is the block $[{\cal M}^{(\nu)}]_{33}\equiv \mu$ in Eq.~\eqref{lowscale}, the model is usually called \emph{inverse Seesaw} \cite{Mohapatra:1986bd}, and the light neutrinos have masses of order $m_\nu \sim \mu\,  m_D^2 / M_R^2$. Instead, if the small parameter is the block $[{\cal M}^{(\nu)}]_{13} \equiv \varepsilon$, the model is called \emph{linear Seesaw}  \cite{Malinsky:2005bi}, because the light neutrino masses are linear in $m_D$, of order $m_\nu \sim \varepsilon\, m_D/M_R$. 

The inclusion of $B-L$ as a gauge symmetry was also studied in the scenario of \emph{inverse seesaw} \cite{Khalil:2010iu}.    In the local $B-L$ scenario of inverse seesaw, in addition to the extra neutrinos $\nu_R$ and $S$ included in the original low scale models of Eq.\,(\ref{lowscale}),  one needs an additional neutrino singlet, $S'$, again in order to have anomaly cancellation; moreover a $Z_2$ parity is assumed, under which $S'$ is odd while all other fermions are even, so that $S'$ does not couple to other fermions \cite{Khalil:2010iu}.  In this model,  a scalar singlet acquires a vacuum expectation value thus breaking $B-L$ spontaneously, and at the same time the small mass parameter of the inverse seesaw  is generated by a non-renormalizable effective operator containing this scalar, which implies the inclusion of an extra scale --the high energy cutoff of this operator. This extra scale summarizes the effect of new underlying physics at low energies, where the theory resembles an inverse seesaw scenario, with a phenomenology that  could be distinguished from the original local $B-L$ model of Ref.~\cite{Khalil:2006yi}. It is also possible to construct a model with purely renormalizable operators, without the need of the explicit inclusion of an effective cutoff scale \cite{Basso:2012ti}. 

In this work we explore yet another scenario with local $B-L$ symmetry, which resembles a \emph{linear seesaw} model in the neutrino sector after spontaneous symmetry breaking. In addition, unlike the work in Ref.~\cite{Khalil:2010iu}  where the small mass term of the seesaw matrix is included as an effective non renormalizable term, here we generate the small mass term through a purely renormalizable interaction with an extra scalar doublet. In such case, the smallness of the masses is not due to the high cutoff scale of an effective operator, but must originate from the pattern of symmetry breaking triggered by the scalar potential. 
In such scenario, the remains of spontaneous symmetry breaking contain a heavy $Z'$ as before, but also a richer spectrum with neutral and charged heavy scalars. One caveat of $B-L$ broken at the electroweak scale is the possibility that any preexisting baryon asymmetry in the early Universe could be washed out by $B-L$ violating processes in equilibrium together with sphaleron processes \cite{Blanchet:2008zg}. In the present study we will assume this is not the case, leaving the detailed study of the conditions for further work. 
In Section II we construct the model with its pattern of symmetry breaking, its particle content and interactions. Sections III is devoted to the phenomenology of the model. Finally, in Section IV we state our summary and conclusions.

\section{A gauged $B-L$ model for Linear seesaw}

\subsection{Fields and charges}

Starting from the neutrino spectrum and mass matrix as in Eq.~(\ref{lowscale}), we want to construct a model whose mass matrix, expresed in a similar basis $(\nu_L^{\ c}, \nu_R, S)$, corresponds to a \emph{linear seesaw} type:
\begin{eqnarray}
{\cal M}^{(\nu)} &=&\BM{ccc} 0 & m_D & \varepsilon 
\\
\vspace{2pt}
	 m_D^T & 0 & M_\chi \\ 
	\varepsilon^T & M_\chi^T & 0\EM .
\label{M_nu_1}
\end{eqnarray}
In the three-family case, each ``element'' shown in Eq.~(\ref{M_nu_1}) is actually a $3\times 3$ block. Solving Eq. (\ref{M_nu_1}) in a block-diagonal form, $\textrm{Diag} (M_{\nu_\ell}, M_{N_1}, M_{N_2})$, the actual neutrino masses are the eigenvalues of the $3\times 3$ blocks:
\begin{eqnarray}
	\label{M_nU'_2_values_m1} M_{\nu_\ell} &=& -\varepsilon (M_\chi^{-1}m_D^T)+h.c. , \\
	\label{M_nU'_2_values_solved_1} M_{N_1} \sim M_{N_2} &\sim & M_\chi+m_D^2M_\chi^{-1}+h.c.
\end{eqnarray}
Here $M_{\nu_\ell}$ is the mass matrix of the light neutrinos observed in oscillation experiments, while $M_{N_1}$ and $M_{N_2}$ correspond to heavier neutral fermions. According to current cosmological constraints, the sum of the light neutrino masses is less than $0.39\, \eV$ \citep{Giusarma:2013pmn}, so  we can expect $ (M_{\nu_\ell})_{ij}\lesssim 10^{-1}\, \eV $.  From Eq.~(\ref{M_nU'_2_values_m1}) we can see that the smallness of $M_{\nu_\ell}$ can arise naturally  \cite{'tHooft:1979bh} from the smallness of $\varepsilon$, a parameter with dimension of mass that softly breaks Lepton Number, without requiring  $M_\chi$ to be much higher than the TeV scale. This is the essence of all low scale seesaw models. In addition, $M_{\nu_\ell}$ is linear in $m_D$, which is proportional to the Standard Yukawa couplings. This is the definition of linear seesaw. 
\bigskip

The neutral lepton mass matrix of Eq.\,(\ref{M_nu_1}) in a local $B-L$ theory is generated,  after the $B-L$ symmetry is spontaneously broken, from the following Yukawa sector:
\begin{eqnarray}
\mathcal{L}_\textrm{\tiny{Yuk}}&\sim &- {y_D}\, \overline{L}\Phi^c\nu_{R}
- {y_M}\, \overline{\nu_{R}^c}\chi S
- {y_\varepsilon}\, \overline{L}H^cS + \ldots
\label{lag_yuk_not_r}
\end{eqnarray}
Here the Yukawa couplings $y_D$, $y_M$ and $y_\varepsilon$ are squared matrices in flavour (family) space, just as the lepton SM doublet $L$ and SM singlets $\nu_R$ and $S$ are row or column vectors in that space. $\Phi$ is the SM Higgs doublet while $\chi$ is an extra scalar singlet and $H$ an extra scalar doublet. 
 Just as in Ref.~\citep{Khalil:2010iu}, since $B-L$ is promoted to be a local symmetry, 
 here one also needs to add an extra neutral fermion $S' $ per family with opposite $B-L$ charge to $S$ 
 in order to cancel the gauge anomaly (our $S$ and $S'$ correspond to $S_2$ and $S_1$ of Ref.~\citep{Khalil:2010iu}, respectively), and a $Z_2$ parity can also be assumed in order to avoid mixings of $S'$ with the other fermions.  
The $S'$ thus remains decoupled from all particles except from $Z'$. In what follows we will not study the phenomenology of the neutral fermions $S'$, leaving the topic for further work, but keeping in mind that $S'$ could contribute to the decay width of $Z'$ as an invisible mode if some $S'$ are light enough. 
The extra scalar fields $\chi$ and $H$, on the other hand,  must be introduced in order to keep the gauge symmetry of the last two terms
in Eq.~\eqref{lag_yuk_not_r}. 
For the last term, which generates the $\varepsilon$ block in the mass matrix, one could prescind from the extra doublet $H$, but that would
require effective non-renormalizable terms that include the standard scalar doublet $\Phi$ and the scalar singlet in a generic form as:
\begin{eqnarray}
 \frac{(\Phi^{\dagger}\Phi)^m\chi^n(\chi^{\dagger})^p}{M_F^{2m+n+p}} \overline{L} \Phi^c S ,
 \label{eff_doublet}
\end{eqnarray}
where $m$, $n$ and $p$ are non negative integers that would depend on the $B-L$ charge assignment of the fields, and the cutoff scale $M_F$  corresponds to the mass of some heavy mediating field. In this work we will not consider such effective scenarios. 

Our construction is a minimal enlargement of the SM gauge group by an extra local $U(1)_{B-L}$ symmetry that generates a linear seesaw model for the neutral leptons, without introducing non-renormalizable effective operators. The linear seesaw model appears when $B-L$ breaks spontaneously by the vacuum expectation values (vev's) of the extra scalars.  We must then assign the appropriate $B-L$ charges to the extra fields, keeping consistency with the known assignment to the SM fields and ensuring that the Yukawa terms are invariant under this $U(1)_{B-L}$ transformation. 
For brevity of notation, in what follows we will designate the $B-L$ charge by the letter $\Upsilon$.   

The $B-L$ charges of the SM fields $L$ (the lepton doublet), $\nu_R$ (the right handed neutrino) and $\Phi$ (the Higgs doublet) are, clearly, 
$\Upsilon_L = \Upsilon_{\nu_R} =-1$, and $\Upsilon_\Phi =0$, respectively. The $B-L$ charges of the extra fields are determined by the $B-L$ invariance of the last two terms in Eq.\,(\ref{lag_yuk_not_r}), 
\begin{equation}
	\label{rules_13_23} \Upsilon_{S_2}-\Upsilon_H=-1 ,\qquad  \Upsilon_{S_2}+\Upsilon_\chi=1,
\end{equation}
and from further restrictions that arise from the scalar potential. 
The most general potential which is invariant under the gauge group  
SU(2)$_\textrm{L}\times$U(1)$_\textrm{Y}\times$U(1)$_\Upsilon$ with the fields $\Phi$, $H$ and $\chi$ is
\begin{eqnarray}
 V_S&=&-\left\{m_1^2(\Phi^{\dagger}\Phi)+m_2^2(H^{\dagger}H)+m_3^2(\chi^{\dagger}\chi)\right\}+ 						\left\{\lambda_1(\Phi^{\dagger}\Phi)^2+\lambda_2(H^{\dagger}H)^2+\lambda_3(\chi^{\dagger}\chi)^2+\lambda_4(\Phi^{\dagger}\Phi)				(H^{\dagger}H)\right.
 \nonumber\\
	 &&\left.+\lambda_5(\Phi^{\dagger}\Phi)(\chi^{\dagger}\chi)+\lambda_6(H^{\dagger}H)(\chi^{\dagger}\chi)+ 							\lambda_7(\Phi^{\dagger}H)(H^{\dagger}\Phi)+\lambda_8[\xi^2(\Phi^\dagger H)+h.c.]\right\}, \quad 
\xi=\chi\textrm{ or }\chi^\dagger  .
\label{pot_V}
\end{eqnarray}
Had we not included the $\lambda_8$-term, the potential would have been invariant under three U(1) transformations, 
$\textrm{U(1)}_\Phi\times \textrm{U(1)}_\textrm{H}\times \textrm{U(1)}_\chi$, instead of the two we need. The extra broken $U(1)$ would generate a physical Goldstone boson, contradicting current phenomenology. The inclusion of the  $\lambda_8$-term eliminates one of the U(1) symmetries and at the same time allows us to identify the other two with U(1)$_\textrm{Y}$ and U(1)$_\Upsilon$. 

The scalar singlet in the $\lambda_8$-term in Eq.~(\ref{pot_V}) is denoted as $\xi$, which alternatively can be chosen as 
$\chi$ or $\chi^\dagger$. 
This is a freedom that we still have and it  affects the $B-L$ charge assignment to the fields.  As it is, the invariance of the  $\lambda_8$-term of Eq. (\ref{pot_V})  imposes the restriction:
\begin{eqnarray}
	 2\Upsilon_\chi\pm\Upsilon_H = 0,  \quad   \textrm{  for $\xi=\chi\textrm{ or }\chi^\dagger$}.
\label{asig_cargas_2}
\end{eqnarray}
The case $\xi=\chi^\dagger$ implies $(\Upsilon_H,\Upsilon_\chi,\Upsilon_{S})=\left(4/3,\  2/3,\ 1/3\right)$, allowing an undesirable Yukawa term in the Lagrangian of the form $\chi^{\dagger}\overline{S^c}S$. This term, after spontaneous symmetry breaking,  would generate a 
$[{\cal M}^{(\nu)}]_{3 3}$ element in the neutrino mass matrix,  and so it is similar to the small element in inverse seesaw scenarios. However, in our scenario this term will not be small, because there is also a Yukawa term of the form $\overline{\nu_{R}^c}\chi S$ that is required to generate the large block $M_\chi$. Consequently, we can only accept the 
$\lambda_8$-term with $\xi \to \chi$:
\[
V_S = \ldots + \lambda_8\  [\chi^2(\Phi^\dagger H) + h.c.].
\]
With this form of the $\lambda_8$ term, the $B-L$ charges are completely defined. Their values are shown in Table \ref{assig_cargas_1}. 
\begin{table}[ht]
	\begin{tabular}{c||c|c|c|c|c|c|c}
		Field & $L$ & $\Phi$ & H & $\chi$ & $\nu_R$ & $S$ & $S'$ \\\hline 
		 SU(2), $Y$, $\Upsilon$ \quad & \  2, -1, -1 \  & \  2, 1, 0 \  & \  2, 1, 4 \  & \  1, 0, -2 \  & \  1, 0, -1 \  & \  1, 0, 3 \  & \  1, 0, -3 \\ 
	\end{tabular}
	\caption{$SU(2)$ representation,  hypercharge $Y$, and $B-L$ charge $\Upsilon$  of the fields in Eq.~(\ref{lag_yuk_not_r}). }
	\label{assig_cargas_1}
\end{table}

Now, with this $B-L$ charge assignment, a $[{\cal M}^{(\nu)}]_{22}$ element in the mass matrix is not forbidden. Indeed, a Yukawa term 
 $y_r\, \overline{\nu_R^c} \chi^\dagger \nu_R$ is allowed, and therefore should be included. After spontaneous symmetry breaking,  $\chi$ gets the v.e.v.  $v_\chi /\sqrt{2}$, generating the block $[{\cal M}^{(\nu)}]_{22} \equiv r=y_r  v_\chi/\sqrt{2}$, and the mass matrix of Eq.~(\ref{M_nu_1}) in the same  basis $(\nu_L^{\ c}, \nu_R, S)$ becomes
\begin{equation}
 {\cal M}^{(\nu)} = \BM{ccc} 0 & m_D & \varepsilon \\ m_D^T & r & M_\chi \\ \varepsilon^T & M_\chi^T & 0\EM .
\label{M_nU'_2}
\end{equation}

After this Yukawa term  is added to  Eq.~(\ref{lag_yuk_not_r}), the full Yukawa sector for the neutral fermions in this model is: 
\begin{eqnarray}
\mathcal{L}_\textrm{\tiny{Yuk}}  &=& - {y_D}\, \overline{L}\Phi^c\nu_{R}
- {y_M}\, \overline{\nu_{R}^c}\chi S
- 											
{y_\varepsilon}\, \overline{L}H^c S 
- \frac{1}{2} {y_r} \,  \overline{\nu_{R}^c} \chi^\dagger \nu_{R} + h.c. 
\label{lag_yuk_con_r} 
\end{eqnarray}

This additional element, however, does not spoil the linear seesaw mechanism: one can show that, when $r\neq 0$ the light neutrino masses are still vanishing with $\varepsilon \to 0$ as in Eq.\,(\ref{M_nU'_2_values_m1}) 
and only the heavy neutrino masses, Eq.~(\ref{M_nU'_2_values_solved_1}), are modified by $r$ in the form
\begin{eqnarray}
	\label{correct_of_m23} M_{N_1} \sim M_{N_2} \sim  M_\chi+m_D^2M_\chi^{-1} + r+h.c. 
\end{eqnarray}

In summary, the corresponding mass eigenstates, considering 3 families, form a set of 9 Majorana neutrinos, three light and six heavy, which we organise as \hbox{$n_\alpha =$ ($\nu_\ell$, $N_1$, $N_2$)}, related to the 9 flavour states \hbox{$\nu_i =$($\nu_L^{\ c}$, $\nu_R$, $S$)}  by a $9\times 9$ mixing matrix $U_{i\alpha}$ as 
\begin{equation}
\nu_i = U_{i\alpha} \ n_\alpha. 
\label{matrixU}
\end{equation}
Notice that in this definition of $U$, all fermion fields are right handed. We use latin indices $i= 1,\ldots, 9$ for the flavour states (the first three are the SM flavours and the other six are the extra neutrinos) and greek indices $\alpha= 1,\ldots, 9$ for the mass states. 
The upper left $3\times 3$ block of $U$ (i.e. for $i,\, \alpha = 1, \, 2,\, 3$) is essentially the well known PMNS matrix, which in orders of magnitude is near unity. The order of magnitude of the other mixing elements $U_{i\alpha}$ per family are schematically given by:
\begin{eqnarray}
\nu_L^{\ c} &\sim& \quad \ \,  \nu_\ell \ +\  b_1 \,  N_1 \, + b_2\,  N_2 ,\nonumber\\
\nu_R &\sim& {\cal O}_\varepsilon \   \nu_\ell \ +\,  a_1  \, N_1 \, +\,  a_2 \, N_2 ,
\label{neutrino_eigen} \\
S &\sim& b_3\    \nu_\ell\ \  + a_3\,   N_1 \ + \, a_4 \, N_2 .\nonumber
\end{eqnarray}
Here the coefficients $a_{1,\ldots 4}$, which relate the extra fields $\nu_R$ and $S$ with the heavy states $N_1$ and $N_2$, are in general of order unity; the heavy-to light mixings  $b_{1,\ldots 3}$ are of order $m_D/M_\chi$,  and finally ${\cal O}_\varepsilon$ is extremely suppressed, of order $\varepsilon/M_\chi$.  Therefore in our model $\nu_R$ and $S$ are mainly admixtures of the heavy neutrinos,
$S$ having a small component of the light states, while the standard $\nu_L$ are mainly admixtures of the light neutrinos, with small -but not necessarily negligible- components of both $N_1$ and $N_2$.
This mixing pattern seems to differ from the \emph{inverse seesaw} version of Khalil \cite{Khalil:2010iu}, where it is claimed that only one of the heavy neutrinos has non-negligible mixing with the standard neutrinos $\nu_L$. 
%

%
\subsection{Symmetry Breaking and the Boson Spectrum}
\label{SSB}

Spontaneous breaking of the local $B-L$ and electroweak symmetry occurs when the 
scalar multiplets $\Phi$, $H$ and  $\chi$ acquire non-zero vacuum expectation values (v.e.v.) at the minimum of the potential shown in Eq.~(\ref{pot_V}). Denoting the field components as 
\begin{eqnarray}
 \Phi =\BM{c} \varphi^+ \\ \frac{1}{\sqrt{2}}(v +\varphi_1+i \varphi_2) \EM,  \quad
 H = \BM{c} \varphi'^+ \\ \frac{1}{\sqrt{2}}( v_H+\varphi'_1+i \varphi'_2) \EM,  \quad 
 \chi=\frac{1}{\sqrt{2}} (v_\chi+\rho_1+i \rho_2),
\label{BSM_scalars}
\end{eqnarray}
the Yukawa sector, Eq.~\eqref{lag_yuk_con_r}, generates the mass matrix of Eq.~(\ref{M_nU'_2}), where 
\begin{eqnarray}
 m_D= \frac{1}{\sqrt{2}}\,   y_D \,  v , 
 \quad 
 M_\chi= \frac{1}{\sqrt{2}}\,   y_M \,  v_\chi,
 \quad 
 \varepsilon=\frac{1}{\sqrt{2}} \,  y_\varepsilon \,  v_H ,
\quad 
r =   \frac{1}{\sqrt{2}}\,   y_r  \,  v_\chi. 
\label{blocks}
\end{eqnarray}
In a natural scenario, we expect the Yukawa couplings, $y_D$, $y_M$, $y_\varepsilon$ and $y_r$, to be just one or two orders of magnitude below unity, so that the rather large spread of scales in the mass matrix should arise mainly through the vev's ($v$, $v_\chi$ and $v_H$).  
First, we have $v\sim 10^2$ GeV (the SM vev.) In addition, we will assume $v_\chi \sim M_\chi$ to be of the order  $10^1\ \TeV$ (this value will be justified below). Finally,  the bound  $m_\nu \lesssim 10^{-1}$ eV on light neutrino masses together with Eqs.~(\ref{M_nU'_2_values_m1}) and (\ref{blocks}) leads to the restriction:
\begin{eqnarray}
 v_H \sim \frac {m_\nu v_\chi}{v} \frac{y_M}{y_D\  y_\varepsilon}  \sim 10^2\ \eV ,
\label{vev_relations}
\end{eqnarray}
where the numerical estimate is given assuming all yukawa couplings of order $\sim 10^{-1}$.  Clearly these v.e.v.s exhibit a wide range of scales. Since they are the field values at the minimum of the scalar potential, Eq.~\eqref{pot_V}, it is important to study whether the potential can accommodate this wide range of scales in a rather natural way and at the same time guaranteeing their stability. The v.e.v.s  satisfy the equations that minimize the potential:
\begin{eqnarray}
0&=&  \lambda_1 v^3 + \left( \frac{\lambda_4+ \lambda_7}{2} v_H^2 
+ \frac{\lambda_5}{2} v_\chi^2 - m_1^2\right) v + \frac{\lambda_8}{2} v_\chi^2 v_H ,
\\
0&=&  \lambda_3 v_\chi^3 + \left( \frac{\lambda_5}{2} v^2 
+ \frac{\lambda_6}{2} v_H^2 {\lambda_8} v  v_H - m_3^2  \right) v_\chi ,
\\
0&=&  \lambda_2 v_H^3 + \left( \frac{\lambda_4+ \lambda_7}{2} v^2 
+ \frac{\lambda_6}{2} v_\chi^2 - m_2^2\right) v_H + \frac{\lambda_8}{2} v_\chi^2 v \label{v_vH_vchi_conditions}.
\end{eqnarray}

We can approximate the solution for $v$ and $v_\chi$ by neglecting all $v_H$ terms in the first two equations, obtaining:

\begin{equation}
v^2 \simeq \frac{2( 2 m_1^2 \lambda_3 - m_3^2 \lambda_5)}{4\lambda_1 \lambda_3 - \lambda_5^2}, \quad 
v_\chi^2 \simeq \frac{2( 2 m_3^2 \lambda_1 - m_1^2 \lambda_5)}{4\lambda_1 \lambda_3 - \lambda_5^2} .
\label{vvx}
\end{equation}

Clearly, we need \emph{some} kind of hierarchy in the quartic coefficients of the potential, $\lambda_i$, in order to get the required v.e.v.'s. One can naturally gather the couplings according to: $\lambda_{1,2,3}$ (quartic term of a single field), $\lambda_6$ (mixing between the new scalars), $\lambda_{4,5,7}$ (mixing between SM-Higgs and new scalars) and $\lambda_8$, the only term not bilinear in every field. 
If the interactions of the SM Higgs with scalars beyond the SM (BSM) are small, we could assume the following hierarchy: 
\begin{equation}
	\lambda_8\ll\lambda_{4,5,7}\ll\lambda_{1,2,3}\sim \lambda_6 \sim 10^{-1}.
\label{hierarchies}	
\end{equation}
In this case, we may assume that $\lambda_5$ is negligible everywhere in Eqs.~\eqref{vvx}, so the latter become:
\begin{equation}
v^2 \sim \frac{m_1^2}{\lambda_1}, \quad v_\chi^2 \sim \frac{m_3^2}{\lambda_3}. 
\end{equation}
Then,  to obtain $v = 246$ GeV and  $v_\chi\sim 10^1$  TeV,  we can choose, e.g., $\lambda_1\sim\lambda_3\sim 10^{-1}$ and
\begin{eqnarray}
\label{cond_for_v_vchi}
	m_1 \sim 10^2\ \GeV, \ \  m_3 \sim 10^4\ \GeV .
\end{eqnarray}

Now, concerning $v_H$, it satisfies Eq.~\eqref{v_vH_vchi_conditions} that can be expressed as:
\begin{equation}
	\label{ec_para_vH}
		\frac{v_H^3}{b} +v_H+c =0,
\end{equation}
with coefficients:
\begin{equation}
b= \frac{(\lambda_4+\lambda_7)v^2+\lambda_6 v_\chi^2-2m_2^2}{2\lambda_2}  ,\quad c=\lambda_8 \frac{ vv_\chi^2}{(\lambda_4+\lambda_7)v^2+\lambda_6 v_\chi^2-2m_2^2}.
\end{equation}
For $b>0$ the equation has only one real solution. Since we want $v_H \ll v$ and we naturally expect $\lambda_8$ very small (as explained below), then if there are no fine tuning cancellations, it is reasonable to expect $b\sim v_\chi^2$ while $c\ll v$. In such case, the real solution is approximately 
\begin{equation}
v_H \approx -c \sim -\frac{\lambda_8 v}{\lambda_6}, 
\end{equation}
which is what we need. Notice that $\lambda_8$ can be negative, as shown below. We do not consider $b<0$, because in that case the real solutions are naturally large. We can obtain the tiny 
v.e.v. $v_H \sim 10^2$ eV consistently with the hierarchy of Eq.~\eqref{hierarchies}, provided
\[
\left| \frac{ \lambda_8 }{\lambda_6} \right| \sim 10^{-9}.
 \]
This parameter can be so small and yet be natural, because an additional $U(1)$ symmetry appears in the scalar potential when $\lambda_8=0$  \cite{'tHooft:1979bh}:  as mentioned below Eq.~\eqref{pot_V}, in such a case the  potential would be invariant under a phase redefinition of the three fields $\Phi$, $H$ and $\chi$,  separately:  $U(1)_\Phi \times U(1)_H \times U(1)_\chi$. When $\lambda_8 \neq 0$, this symmetry reduces to $ U(1)_Y \times U(1)_\Upsilon$.
An additional issue with the scalar potential, Eq.\ (\ref{pot_V}), is to ensure that it is bounded from below in all directions in field space. This is guaranteed if all quartic terms are of a bi-quadratic form $\lambda_{ab}s_a^2s_b^2$ (where $\lambda_{ab}$ denotes the coupling and $s_{a}$, $s_b$ the scalar fields) and all $\lambda_{ab}$ are all positive \citep{Kannike:2012pe}. Consequently we fix $\lambda_{1,\cdots,7}>0$, but we also have the term $\sim \lambda_8 v_\chi^2\, v\, v_H$ which is not bi-quadratic and therefore could be problematic. However, gathering the terms in $\lambda_5$, $\lambda_6$ and $\lambda_8$, and factoring out the magnitude of $\langle \chi \rangle_0 = v_\chi$, one can show that these three terms together are positive definite for all values of the fields, provided 
\[
\lambda_8 ^2 < 4 \lambda_5 \lambda_6,
\]
a condition that is clearly fulfilled according to Eq.~\eqref{hierarchies}. Now, the scalar mass spectrum can be obtained from the diagonalization of the mass matrix 
\[
M_{ij}^2=\left.\frac{\partial^2 V}{\partial \phi_i \partial \phi_j}\right|_{\textrm{vac}}.
\]

As the gauge group $SU(2)_L\times U(1)_Y \times U(1)_{B-L}$ breaks down to $U(1)_{e.m.}$, from 10 degrees of freedom of the scalar sector, four are would-be Goldstone bosons  absorbed by the vector fields $W^\pm$, $Z^0$ and $Z^\prime$, while two charged scalars and four neutral scalars  remain in the physical spectrum as massive particles. The charged scalars, which we call $H^\pm$, are essentially the charged components of the extra doublet denoted in Eq.~\eqref{BSM_scalars} as $\varphi^{\prime \pm}$, except for a tiny admixture of 
the standard $\varphi^\pm$. The $H^\pm$ are expected to be heavy, with masses:
\[
m^2_{H^\pm} \simeq \lambda_6 v_\chi^2 \sim (1 - 10^1\ \TeV)^2,
\]
possibly out of reach for production in current experiments. Of the neutral scalar eigenstates, two of them are mixtures (albeit tiny mixtures) of the standard $\varphi_1$ and the extra $\rho_1$ [c.f. Eq.~\eqref{BSM_scalars}]. We call these fields $h$ and $H_0$:
\begin{eqnarray}
h &=& \varphi_1 \cos\theta_H - \rho_1 \sin\theta_H ,\nonumber\\
\label{h_H0_mass-states} H_0 &=& \varphi_1 \sin\theta_H + \rho_1 \cos\theta_H,
\end{eqnarray}
which are essentially $\varphi_1$ and $\rho_1$, because $\theta_H\simeq \lambda_5 v /(2\lambda_3 v_\chi)$ is very small. 
The masses of these two fields are, respectively:
\begin{eqnarray}
m_{h}^2 &\simeq&  2 \lambda_1 v^2  \ \sim (10^2\ \GeV)^2 ,\\
\label{m_h_H0} m_{H_0}^2 &\simeq&  2 \lambda_3 v_\chi^2 \ \sim (1 - 10^1\  \TeV)^2 .
\end{eqnarray}
The light scalar  $h$ can then be identified with the 125 GeV Higgs particle observed at the LHC, while $H_0$  is  much heavier.
There are  other two massive neutral scalars, $H'_0$ and $H''_0$, which turn out to be essentially the neutral components of the extra doublet, $\varphi'_1$ and $\varphi'_2$ (see Eq.~\eqref{BSM_scalars}), both also  heavy and comparable to $m_{H^\pm}$:
\begin{equation}
\label{m_H'_H''} m_{H'_0}^2 = m_{H''_0}^2 \simeq \lambda_6 v_\chi^2\sim (1 - 10^1 \ \TeV)^2.
\end{equation}
Consequently, with our numerical choices of v.e.v.'s, we obtain one neutral scalar with the mass of the SM Higgs (proportional to $v\simeq 246$ GeV), and three other neutral scalars and two charged scalars, all with masses near $10^1\, \TeV$ (proportional to $v_\chi$). Notice that  no scalar masses are proportional to the light $v_H$, due to the form of the scalar potential. 

Now, concerning the gauge bosons associated to the broken generators of the gauge group, they get masses through the Higgs mechanism, absorbing the would-be Goldstone bosons into their longitudinal polarizations. Their masses arise from the kinetic terms in the Lagrangian for the scalar fields:
\begin{eqnarray}
\label{L_kin}
\mathcal{L}_\textrm{kin}^{(s)} =\left(D_{\mu} \Phi\right)^\dagger(D^\mu \Phi) +
\left(D_{\mu} H \right)^\dagger(D^\mu H) +
\left(D_{\mu} \chi\right)^\dagger(D^\mu \chi) + \ldots
\end{eqnarray}
where the covariant derivative is, in general for any of our scalar mutiplets:
\begin{equation}
D_\mu = \partial_\mu + i g_2 t_a W^a_\mu + i g_1 \frac{Y}{2} B_\mu + i g'_1 {\Upsilon} B^\prime_\mu .
\label{cov_D}
\end{equation}
Here $t_a=\sigma_a/2$ ($a=$1, 2, 3) are the SU(2) generators in the doublet representation, and the values of the charges $Y$ and $\Upsilon$ for the different fields are shown in Table \ref{assig_cargas_1}.
 After spontaneous symmetry breaking, the terms in $\mathcal{L}_\textrm{kin}^{(s)}$ quadratic in the vector fields become a mass matrix for these fields. The charged vector fields, $W^\pm_\mu$,  acquire  a mass
\begin{eqnarray}
	\label{chgb_mass} M_W^2&=& \left.M_{W}^2\right|_\textrm{{SM}}\left[1+\left(\frac{v_H}{v}\right)^2\right] ,
\end{eqnarray}
which is practically the same as in the SM. On the other hand, the neutral vector fields $W^3_\mu $, $B_\mu$ and $B^\prime_\mu$ have a mass matrix
\begin{eqnarray}
{\cal M}_0^2 &=& \frac{1}{4}
\BM{ccc} 
g_2^2(v^2+v_H^2)       &    -g_2g_1(v^2+v_H^2)    &    -8g_2g'_1v_H^2 \\ 
-g_2g_1(v^2+v_H^2)     &      g_1^2(v^2+v_H^2)    &     8g_1g'_1v_H^2 \\ 
-8g_2g_1'v_H^2         &     8g_1g'_1v_H^2        &   {16g_1'^2(v_\chi^2+4v_H^2)}
\EM ,
\label{vectormass}
\end{eqnarray}
which gives the following mass eigenvalues:
\begin{eqnarray}
\nonumber M_A &=& 0 \, ,\\
\label{gauge_bosons_masses} M_Z^2 &\simeq &\left.M_Z^2\right|_\textrm{{SM}}\left[1-\frac{16 g_1'^2}{g_1^2+g_2^2} \left(\frac{v_Hv_\chi}			{v^2}\right)^2+ \left(\frac{v_H}{v}\right)^2\right]\, , \hspace{20mm} \\	
\nonumber M_{Z'}^2 &\simeq & 4 g_1'^2v_\chi^2
\end{eqnarray}
where the photon, $A$, is massless and the $Z$ boson has practically the same mass as in the SM, while the extra neutral vector boson $Z'$ is practically $B'$, due to the smallness of $v_H$ (compared to $v$ and $v_\chi$) on the third row and column in Eq.~\eqref{vectormass}.  $M_{Z'}$ has an experimental lower bound of around 600 GeV from experiments at the Tevatron, and there is also a lower bound on the ratio $M_{Z'}/g'_1$ set by LEP II experiments 
\cite{Carena:2004xs}, 
which gives a direct bound on our extra v.e.v. $v_\chi$, namely 
\begin{equation}
	\label{constraint_for_v_chi}
	\frac{M_{Z'}}{2 g'_1}= v_\chi>3 \ \TeV, 
\end{equation}
and an updated analysis \cite{Heeck:2014zfa} of the LEP II results \cite{Schael:2013ita} claims the slightly higher bound of  $3.4\ \TeV$ at 95\% C.L. 

We notice that the $\rho$ parameter in this model acquires a very small deviation with respect to the SM value: 
\begin{eqnarray}
	\rho &\equiv& \frac{M_W^2}{M_Z^2\cos^2\theta_W}\simeq \rho_\textrm{SM} +\frac{16 g_1'^2}{g_1^2+g_2^2}\left(\frac{v_Hv_\chi}{v^2}\right)^2 .
\end{eqnarray}
To be consistent with experiment, this correction should be smaller than the current uncertainty on $\rho$ coming from the values of $M_W$, $M_Z$ and $\cos\theta_W$, which  is $\Delta\rho \sim 10^{-4}$ \cite{Beringer:1900zz}. Using Eq.~\eqref{vev_relations} for $v_H$, the current bound $m_\nu \gtrsim 0.05$ eV from atmospheric neutrino oscillations, and assuming all yukawas and $g_1'$ of order $10^{-1}$, we find the upper bound  $ v_\chi \lesssim 10^5 \ \TeV$,  which is clearly above the scale of any \emph{low scale seesaw}.  

Concerning the lepton interactions, those with the scalar fields are determined by the Yukawa sector, Eq.~\eqref{lag_yuk_con_r}, while those with the massive vector fields are determined by the lepton kinetic sector:
\begin{equation}
{\cal L}_{kin}^{(l)}  = i\, \overline L  \gamma^\mu D_\mu L + i\, \overline \ell_R \gamma^\mu D_\mu \ell_R + i\, \overline \nu_R \gamma^\mu D_\mu \nu_R + i\, \overline S \gamma^\mu D_\mu S+ \ldots
\end{equation}
The covariant derivative in the first term has exactly the form in Eq.~\eqref{cov_D} with the $Y$ and $\Upsilon$ charges
corresponding to the SM lepton doublet $L = ( \nu_L, \ell_L)^T$ (see Table \ref{assig_cargas_1}), while in the last two covariant derivatives only the $B'_\mu$ field is present because $\nu_R$ and $S$ are sterile under the SM interactions. Then, according to Eq.~\eqref{matrixU}, the couplings of the mass eigenstate neutrinos $n_\alpha$ to the vector bosons are:

\begin{eqnarray}
	\mathcal{L}_{W} &=& \frac{g_2}{\sqrt{2}}W_\mu^-\sum_{i=1}^{3}\sum_{\alpha=1}^{9}B_{i \alpha}\ \overline{\ell_i}\gamma^\mu P_L\, n_\alpha+h.c. \,  , \quad B_{i\alpha} \equiv \sum_{k=1}^3 V_{ik} U_{k\alpha}^\ast ,
\label{NtoW}	
\\
\mathcal{L}_{Z} &=& \frac{g_2}{2\cos\theta_W}Z_\mu \sum_{\alpha =1}^{9}  \sum_{\beta=1}^{9} C_{\alpha\beta}\  
	\overline{n}_\alpha \gamma^\mu P_R \, n_\beta  , \quad C_{\alpha\beta} \equiv \sum_{i=1}^{3} U_{i\alpha}^\ast U_{i\beta} ,
\label{NtoZ}	
\\
\mathcal{L}_{Z' } &=& - g'_1  Z'_\mu \sum_{\alpha =1}^{9}  \sum_{\beta=1}^{9} C'_{\alpha\beta}\  
	\overline{n}_\alpha \gamma^\mu P_R \, n_\beta  , \quad C'_{\alpha\beta} \equiv  (U^\dagger \tilde\Upsilon U)_{\alpha\beta},
\label{NtoZp}	
\end{eqnarray}
where $P_{L,R}$ are the chiral projectors. In the first line, we included $V_{ik}$ as the matrix that relates the mass to the flavour states of the charged leptons, although it can be taken to be the unit matrix by definition.  In the last expression,   $\tilde\Upsilon= \textrm{Diag}( I, - I,  3\ I)$, where $I$ is the $3\times 3$ unit matrix. These diagonal elements correspond to the $B-L$ charges  of $\nu_L^{\ c}$, $\nu_R$ and $S$, respectively. Here it is important to remember that $n_\alpha$ are Majorana neutrino fields, so that 
$\overline n_\alpha \gamma^\mu P_R n_\beta =  - \overline n_\beta \gamma^\mu P_L n_\alpha$.

\section{Phenomenology.}

As described in the previous section, this model contains several particles beyond the SM:
besides the light neutrinos $\nu_\ell$, which are Majorana fields, the spectrum includes two heavy Majorana neutrinos per family, $N_1$ and $N_2$, one heavy neutral vector boson $Z'$, a pair of heavy charged scalars $H^\pm$, and four neutral scalars: $h$ (with $m_h \sim 125$~GeV) and three heavier, $H_0$, $H'_0$ and $H''_0$.   The effect of these heavy fields
may show by direct production at high energies or as virtual states that induce rare processes at low energies, in particular processes that exhibit lepton flavour violation (LFV). We first study briefly these LFV processes, which are practically the same as in the inverse seesaw model. As such, more than finding distinctive signals of the model in LFV processes, we study them for consistency, because they can impose bounds on the model parameters. We then proceed to analyse the direct production of the heavy particles at high energies.

\subsection{Lepton flavor violating processes}

\begin{figure}[h!]
	{\includegraphics[width=8cm,height=5cm]{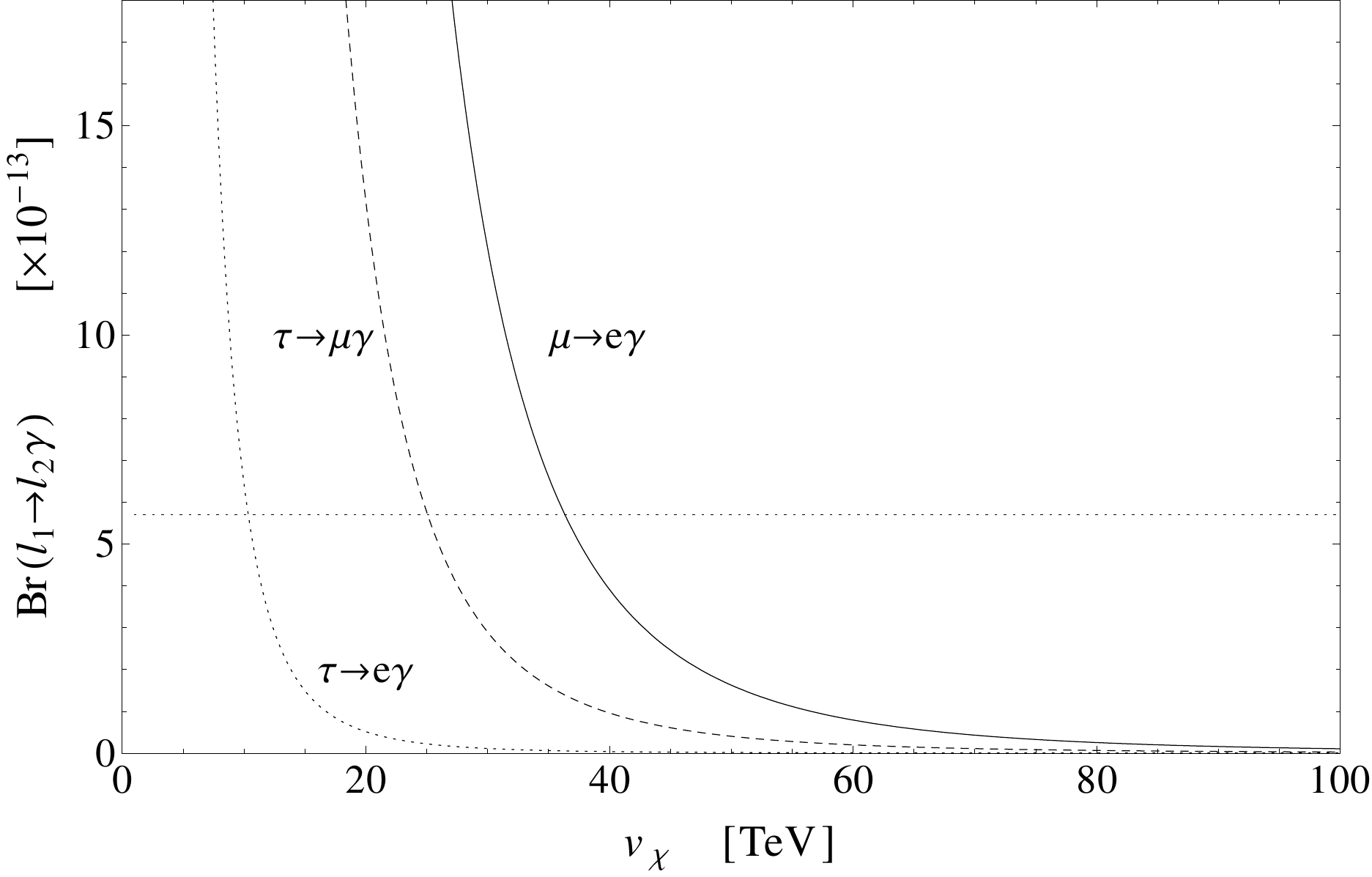}}
\caption{Branching Ratios for the lepton flavour violating decays  $\mu\rightarrow e\gamma$ (solid line),  $\tau\rightarrow e\gamma$ 
(dotted line) and $\tau\rightarrow\mu\gamma$ (dashed line) as a function of the v.e.v. $v_\chi$, assuming all Yukawa couplings to be of order unity. The horizontal line corresponds to the current experimental upper bound on $\mu\to e\gamma$, while the upper bounds on the other processes are horizontal lines five orders of magnitude above (not shown). The analytic expressions for the rates in terms of masses and mixings were taken from Ref.~\citep{Ilakovac:1994kj}. 
}
\label{fig_BR_Ilakovac}
\end{figure}

The neutrino sector of both inverse and linear seesaw models are similar, so the types of LFV processes that appear in both models are also similar. These processes have been studied elsewhere, however  we should review them here because their current upper bounds may already impose restrictions on our model parameters. As shown in  Table~\ref{thebounds}, the strongest bound comes from $\mu\to e\gamma$. The contributions to this process start at 1 loop level, and are sensitive to the masses of the heavy neutrinos in the loop, which are $m_N \propto v_\chi$, as well as the mixings $B_{\mu {N_i}}$ and $B_{e {N_i}}$, which are of order $B_{\ell N} \sim v/v_\chi$. 
Taking the yukawa couplings to be of order unity, the bound on a LFV process translates into a bound for $v_\chi$, as shown in Fig.~\ref{fig_BR_Ilakovac}. 
\begin{table}[h!]
	\begin{tabular}{|c|c|}
	\hline
		Process & Branching ratio  \\ \hline\hline
		$\mu\rightarrow e\gamma$ & $<5.7\times 10^{-13}$  \\\hline
		$\tau\rightarrow e\gamma$ & $<3.3\times 10^{-8}$  \\\hline
		$\tau\rightarrow\mu\gamma$ & $<4.4\times 10^{-8}$  \\\hline
	\end{tabular}
	\caption{Current upper bounds on the branching ratios of the indicated LFV processes. The bound on $\mu\to e\gamma$ is from Ref.~\cite{Adam:2013mnn}, while the rest are from Ref.~\cite{Beringer:1900zz}.  } 
	\label{thebounds}
\end{table} 
Comparing this figure with the values in Table \ref{thebounds} we see that the LEP II bound for $v_\chi$ in Eq.\,(\ref{constraint_for_v_chi}) is stronger than those obtained from all LFV processes except $\mu\to e\gamma$, which provides a stronger bound $v_\chi\gtrsim 35 \, \TeV$ when the yukawas are of order unity.  For smaller yukawa values, the bound on $v_\chi$ will be correspondingly lower.

We also studied the non-radiative LFV processes $\tau\to e e e$, $\tau\to \mu e e$, $\tau\to\mu\mu e$ and $\tau\to\mu \mu \mu$, as functions of $v_\chi$. Even for yukawa couplings near unity, our model gives considerably smaller values for these branching ratios, namely $10^{-13}$ or less, for 
$v_\chi \gtrsim 10$ TeV, while the current experimental bounds for the branching ratios of these processes are around $10^{-8}$. Consequently these latter processes  do not provide bounds on the model, unlike the decays shown in Table~\ref{thebounds}.

As a final point in this section, let us briefly comment on the effect of scalars on the LFV processes. As previously mentioned, this linear seesaw model contains heavy charged scalars as well. These particles also enter in the loop of the LFV processes (the diagrams can be drawn by substituting the scalars for the $W$ bosons). However, due to their larger masses, their contributions are subdominant, and so they do not significantly affect the LFV rates.  In what follows, we will focus our study on the production of the heavy particles in high energy processes. 

%


%
%

%

\subsection{Heavy Neutrinos and $Z'$}

Let us now address the heavy neutrinos, which in our model should have masses comparable to $v_\chi$, i.e. $1 - 10\, \TeV$, so in principle they could be produced at the LHC.
The main production channels of heavy neutrinos at a high energy collider are through decays of virtual $W\to \ell N_i$, virtual $Z\to \nu_\ell N_i$ or $N_i N_j$ and either real or virtual $Z' \to \nu_\ell N_i$ or $N_i N_j$. The corresponding couplings can be extracted from Eqs.~\eqref{NtoW}, \eqref{NtoZ} and \eqref{NtoZp}, respectively. The largest production rates clearly occur if real $Z'$ are produced and subsequently decay into heavy neutrinos. The neutrino production rates will then be proportional to the partial decay rates of $Z'$:
\begin{eqnarray}
\Gamma(Z'\to N_i N_i) &=& \frac{{g_1'}^2}{24 \pi} C_{N_i N_i }^{\prime \, 2} M_{Z'} \left( 1- \frac{4 m_N^2}{M_{Z'}^2}\right)^{3/2} ,
\\
\Gamma(Z'\to N_1 N_2) &\sim& \frac{{g_1'}^2}{12 \pi} M_{Z'} \left\{ 
(\textrm{Re } C'_{N_1 N_2 })^2  \left( 1- \frac{4 m_N^2}{M_{Z'}^2} \right) + 
(\textrm{Im } C'_{N_1 N_2 })^2  \left( 1+ \frac{2 m_N^2}{M_{Z'}^2} \right)
\right\}
\sqrt{ 1- \frac{4 m_N^2}{M_{Z'}^2} } ,
\\
\Gamma(Z'\to \nu_\ell  N) &= & \frac{{g_1 '}^2}{12 \pi} |C'_{\nu_\ell N}|^2 M_{Z'} \left( 1- \frac{m_N^2}{M_{Z'}^2}\right) 
\left( 1 - \frac{m_N^2}{2 M_{Z'}^2}  - \frac{m_N^4}{2 M_{Z'}^4} \right) ,
\end{eqnarray} 
where the mixing coefficients $C'_{\alpha\beta} (= C_{\beta\alpha}^{\prime \ast})$  are given in Eq.~\eqref{NtoZp}; in the second expression, for simplicity we have assumed all heavy neutrino masses to be similar, denoted as $m_N$. According to Eqs.~\eqref{neutrino_eigen}  and \eqref{NtoZp},  $C'_{NN} \sim {\cal O}(1)$, while $C'_{\nu_\ell N} \sim {\cal O}(m_D/M_\chi) \sim v/v_\chi$ are suppressed. Consequently, the main production mode of heavy neutral leptons is  through real decay of $Z'$ into a heavy pair, $N N$, as long as $M_{Z'}> 2 m_N$. 

The actual value of $C' _{NN}$, for a given family, may vary between $1$ and $3$ depending on the mixing of the two extra states $\nu_R$ and $S$ into $N_1$ and $N_2$. This can be easily seen from the definition of $C'_{\alpha \beta}$ in Eq.~\eqref{NtoZp}. Let us consider just one family, for simplicity. Neglecting the heavy-to-light elements $U_{\nu_L N_1}$ and $U_{\nu_L N_2}$, and considering a generic mixing angle $\theta$ between  the second and third components, namely $U_{\nu_R N_1 }= U_{S N_2} = \cos\theta$ and $U_{\nu_R N_2}= - U_{S N_1}=\sin\theta$, we will have
\begin{eqnarray}
C'_{N_1 N_1} &=&  - \cos^2\theta + 3 \sin^2\theta , 
\nonumber \\
C'_{N_1 N_2} &=&  (-1-3) \cos\theta \sin\theta , 
\label{rotations} \\
C'_{N_2 N_2} &=&  - \sin^2\theta + 3 \cos^2\theta .
\nonumber
\end{eqnarray}
Accordingly, if $N_1$ and $N_2$ have comparable masses, within this model the production rate of the pairs $N_1 N_1$ and $N_2 N_2$ could be  equal  ($\theta = \pi/4$), they could differ by a factor $9$ ($\theta\to 0$ or $\pi/2$), or it may even occur that one of these channels tends to vanish ($\theta\to \pi/6$ or $\pi/3$). However, after we analysed numerically the neutrino mass matrix, Eq.~\eqref{M_nU'_2}, the mixing depends mainly on the submatrices $M_\chi$ and $r$ and, without assuming further restrictions, one should expect both channels to be of similar size, i.e. $\theta$ is unlikely to be zero or $\pi/2$.

\begin{figure}[h!]
	\subfigure[]{\includegraphics[width=8cm,height=5cm]{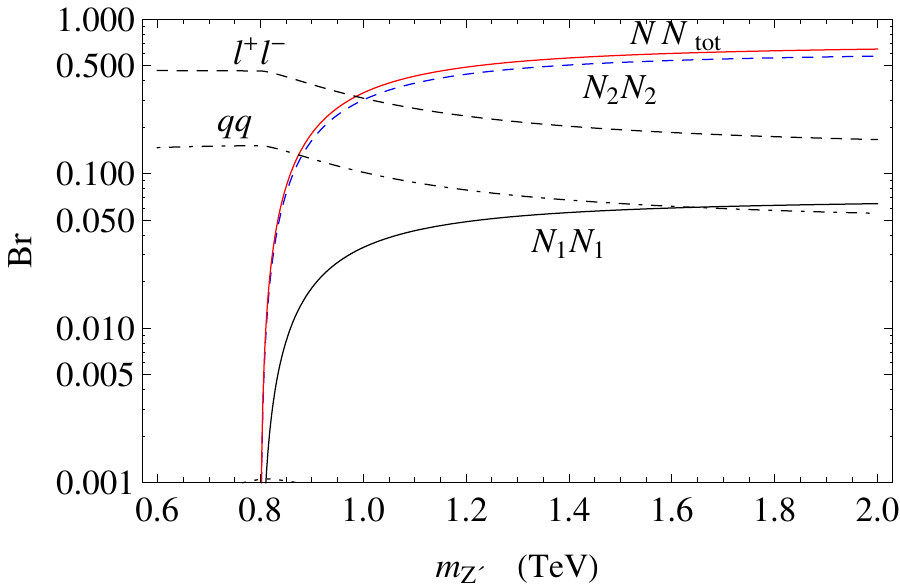}}\hspace{1cm}
	\subfigure[]{\includegraphics[width=8cm,height=5cm]{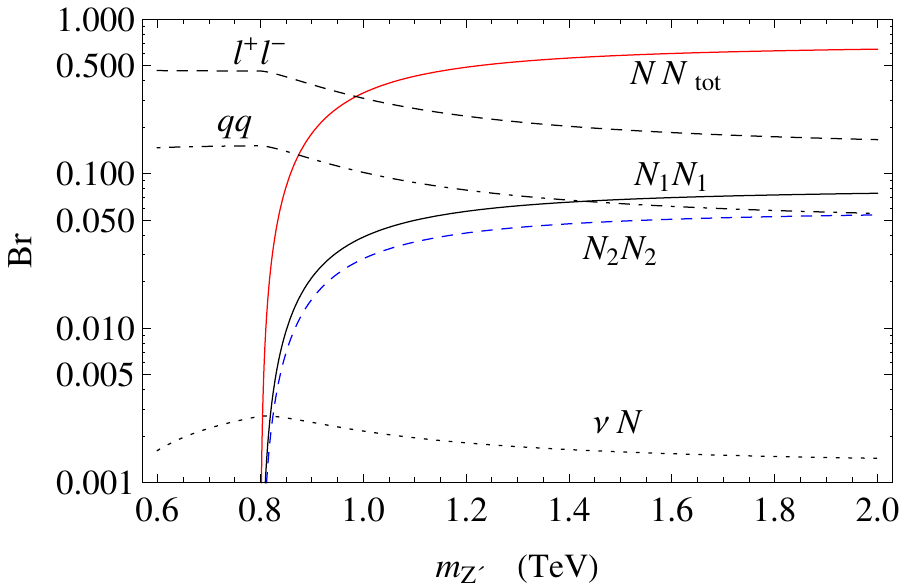}}
	\caption{ Branching ratios for $Z'$ to fermion pairs, considering one family only.  For three families, the lepton pair line $\ell^+\ell^-$  represents the sum $e^+e^- + \mu^+\mu^- +\tau^+\tau^-$ (all three are equal), and the quark pair line $q \bar q$ represents the sum of 
three quark flavors. The heavy neutrinos are assumed to have $m_N = 400$ GeV in order to exhibit the threshold for $NN$ production, and the heavy-to light suppression $m_D/M_\chi = 0.03$, which controls the $\nu_\ell N$ production. The Figures differ in the mixing of the two heavy neutrinos: (a) $\theta =0$, (b) $\theta= 46^\circ$. $NN_{tot}$ means $N_1 N_1 + N_2 N_2 + N_1 N_2$. The mode $N_1N_2$ vanishes in (a) and is comparable to $N_1N_1$ or $N_2N_2$ in (b), but it is not explicitly shown in the figure. The mode $Z'\to S' S'$ was disregarded in the figure, which would contribute to the $Z'$ width only in the case $M_{Z'}> 2 m_{S'}$.}
	\label{fig_BrZp}
\end{figure}

In Fig.~\ref{fig_BrZp} we exemplify the production of heavy neutrinos through $Z'$ decay, for two different mixings $\theta$. Comparing the two figures one can notice that the production of all $NN$ pairs does not vary much, but the proportion of $N_1 N_1$ vs. $N_2 N_2$ does, keeping in mind that the experimental distinction between $N_1$ and $N_2$ could be done only if they differ in mass. 

Also shown in the figures are the $Z'$ decays into SM fermion pairs. The expression for the decay into a charged lepton pair is:
\begin{equation}
\Gamma(Z'\to \ell^+ \ell^-) = \frac{{g_1'}^2}{12 \pi} M_{Z'}  
 \left( 1+ \frac{2 m_\ell^2}{M_{Z'}^2} \right)
\sqrt{ 1- \frac{4 m_\ell^2}{M_{Z'}^2} } ,
\end{equation}
and a similar expression for $Z'\to q \bar q$ with a extra factor $1/3$ (a factor 3 for colour and $(1/3)^2$ for $B-L$, respectively). 
As shown in the figure, these modes have smaller but comparable rates to those of $Z'\to NN$, if the latter are kinematically allowed.

Now, if $N N$ production through $Z'$ decay is kinematically forbidden, the next channel to use is $Z'\to \nu_\ell N$, provided $m_N < M_{Z'} < 2 m_N$.  However, in all possible scenarios this channel is suppressed by heavy-to-light mixing $|C_{\nu N}|^2$ (which is $10^{-2}$ or much less), so $N$ will be more difficult to observe. 

To finish our discussion on $N$ production, if $M_{Z'}$ turns out to be smaller than any of the heavy neutrinos, then the  production of $N_i$ will occur only through virtual $W$, $Z$ or $Z'$ at much smaller rates. In such case, the main signal of this model will not be the heavy neutrinos but the $Z'$, a signal which is common to all local $B-L$ models. 

Let us then complement our discussion on the production and decay of $Z'$ within this model. As shown in the previous section, $Z'$  practically does not mix with the standard $Z^0$ boson, and so it couples through $B-L$ charge  with equal strength to all SM fermions; therefore it is \emph{produced} just as in the \emph{inverse seesaw} model of Ref.~\cite{Khalil:2010iu}: in proton-proton collisions, they are produced through quark-antiquark annihilation $u\bar u, d\bar d\to Z'$, or $Z'$-strahlung $q\to q Z'$.  

Now, considering the decays of $Z'$, as described above the branching ratios will depend on whether the $Z'\to NN$ modes are kinematically allowed, and if they are, they will also depend on the mixing between the heavy neutrinos, as shown in Fig.~\ref{fig_BrZp}, but in general these modes will dominate the $Z'$ decay. On the other hand, if the $Z'\to NN$ modes are not allowed, then the main decays of $Z'$ will be into SM fermion pairs:  $Z'\to \ell^+ \ell^-$, $\nu_\ell \nu_\ell$ and  $q\bar q$, a common feature with the model of Ref.~\cite{Khalil:2010iu}. Since the $B-L$ charges of all fermions are of order unity, all these branching ratios are sizeable.

\subsection{Scalars}

\begin{figure}[h!]
 {\includegraphics[width= 0.9\textwidth]{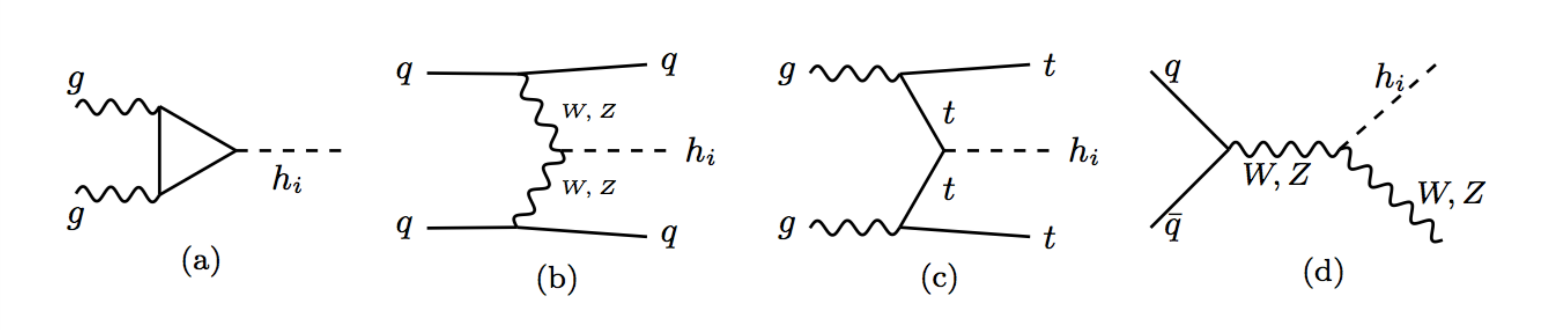}}
	\caption{Production modes of the SM Higgs, also applicable to the extra scalar $H_0$.}
	\label{Higgs_SM}

\end{figure}

Let us now consider the scalars in the model.   Our light Higgs $h$, according to Eq.~\eqref{h_H0_mass-states} is practically the SM Higgs with a small admixture ($\theta_H \lesssim 10^{-3}$) of the scalar singlet, therefore the production and decay of $h$ is just as in the SM. Concerning the extra scalars, their production at a hadron collider can be studied in comparison to the production modes of the SM Higgs at the LHC, which are shown in Fig.~\ref{Higgs_SM}.

On the one hand, the heavy neutral $H_0$, according to Eq.~\eqref{h_H0_mass-states}, is essentially the scalar singlet with a small 
admixture $\theta_H$ of the SM Higgs. Since the scalar singlet does not couple either to  $W^{\pm}$, $Z^0$ or to quarks (because it is a singlet and because of its $B-L$ charge, respectively),  $H_0$ couples to these fields only through its small admixture with the SM Higgs: 
\[
g_{H_0 q \bar q } = \theta_H \frac{m_q}{v} , \quad  g_{H_0 W^+ W^-} = \theta_H\  g_2 M_W , \quad g_{H_0 Z Z}  = \theta_H\  \frac {g_2  M_Z}{\cos\theta_W},
\]
all of which are suppressed by $\theta_H \sim 10^{-3}$ or less. Consequently the production of $H_0$ is suppressed with respect to all the production modes of the light scalar $h$ shown in Fig.~\ref{Higgs_SM} in the same proportion $\theta_H$, with the sole exception of a diagram similar to Fig.~\ref{Higgs_SM}.b, where $H_0$ receives a contribution from $Z' Z'$ fusion, which is not suppressed. Indeed, the couplings involved in this contribution are:
\[
g_{H_0 Z'Z'}=4 g_1' M_{Z'}, \quad  g_{Z' q \bar q} = -\frac{g_1 ' }{3} .
\]
These couplings are not suppressed because $Z'$ is the $B-L$ (broken) gauge field, and so it couples directly to fermions and to the extra scalar multiplets, which  all have $B-L$ charge. Consequently, in a collider with enough energy, the production of the extra scalar $H_0$ will show an enhancement in the mode $q q \to q q H_0$ due to $Z' Z'$ fusion, as compared to the production modes of the standard scalar $h$.

\begin{figure}[h!]
 {\includegraphics[width= 0.5\textwidth]{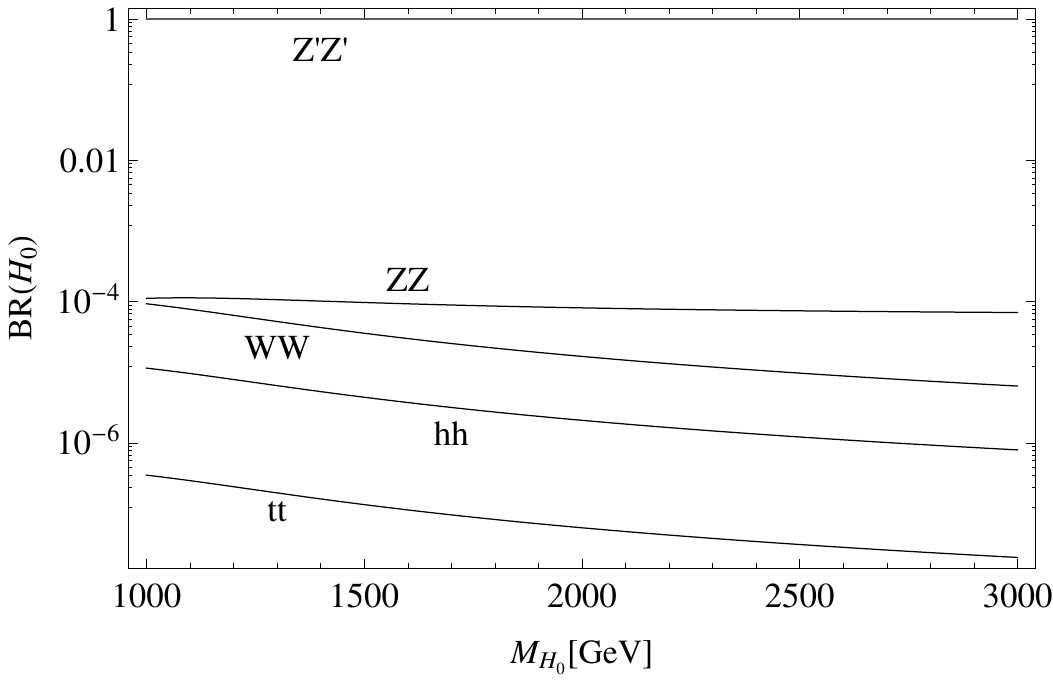}}
	\caption{The main branching ratios for $H_0$ decays, assuming $m_{H_0}> 2 M_{Z'}$, for the benchmark parameters 
	$\lambda_5/\lambda_3 = 10^{-2}$ and $v/v_\chi = 10^{-1}$. The $t\bar t$ channel is actually representative of any $q\bar q$ pair. In this comparison, the channels $H_0 \to N_iN_j $, which depend on other couplings, are assumed to be kinematically forbidden. }
	\label{fig_BR_N_H_zero}
\end{figure}

Concerning the decays of $H_0$, the main channels are into boson pairs, with rates given by:
\begin{eqnarray}
\Gamma(H_0\to Z'Z')
&\simeq & 
\frac{\lambda_3}{16\pi} m_{H_0} \left[1-\frac{4M_{Z'}^2}{m_{H_0}^2}\left(1-\frac{3M_{Z'}^2}{m_{H_0}^2}\right)\right]\sqrt{1-\frac{4M_{Z'}^2}{m_{H_0}^2}} ,
\\
\Gamma(H_0\to W^+W^-)&\simeq & \frac{1}{32\pi} \frac{\lambda_5^2}{\lambda_3}m_{H_0} \sqrt{1-\frac{4M_W^2}{m_{H_0}^2}} ,
\\
\Gamma(H_0\to ZZ)&\simeq & \frac{\cos^4\theta_W}{8\pi} \frac{\lambda_5^2}{\lambda_3} m_{H_0}   \left(1-\frac{4M_Z^2}{m_{H_0}^2}\right)^{3/2}  ,
\\
\Gamma(H_0\to hh)
&\simeq & 
\frac{1}{256\pi}\frac{\lambda_5^2}{\lambda_3}m_{H_0}\sqrt{1-\frac{4m_h^2}{m_{H_0}^2}}  ,
\end{eqnarray}

Accordingly, $H_0$ will decay mainly into $Z'Z'$, provided $m_{H_0}$ is above this threshold. The next dominant channels are into pairs of SM bosons, with a suppression $(\lambda_5/\lambda_3)^2 $ with respect to the former channel. In principle this parameter is unknown, but in this model we estimate it to be small ($\lesssim 10^{-3}$) so that the scalar potential can reproduce a reasonable \emph{low scale seesaw} pattern in the neutrino sector (see Eq.~\eqref{vvx} and related paragraphs). 

Fig.~\ref{fig_BR_N_H_zero} shows the corresponding branching ratios of the $H_0$ decays, assuming the $Z' Z'$ channel is kinematically allowed. The figure also includes the $t\bar t$ channel. For simplicity, in this figure the decays into heavy neutrino pairs are supposed to be kinematically forbidden. If, on the contrary, these modes were allowed, the $H_0$ decays into fermion pairs will be dominated by heavy neutrinos, because the decays into SM fermion pairs are relatively suppressed: $H_0$ couples to SM fermions only due to its small admixture $\theta_H$ of the SM Higgs, which in turn couples very weakly to fermions except the top quark. The corresponding expressions are:

\begin{eqnarray}
\nonumber 
\Gamma(H_0\to N_1N_2)&\simeq&
\frac{1}{16\pi}
m_{H_0} \left\{ 
 |g_N|^2   \left(1-\frac{2 m_{N}^2}{m_{H_0}^2}\right) 
- 2 \textrm{ Re} \left[g_N^2\right] 
\frac{m_N^2 }{m_{H_0}^2} 
\right\} 
\sqrt{1-\frac{4 m_{N}^2}{m_{H_0}^2} } ,
\\
\label{dec_H0->ni_nj-particulares}  
\Gamma(H_0\to N\nu_\ell)&=&\frac{1}{16\pi}|g_{\ell }|^2 m_{H_0}\left(1-\frac{m_{N}^2}{m_{H_0}^2}\right)^2 ,
\\
\nonumber 
\Gamma(H_0\to ff)
& \simeq &
\theta_H^2  \frac{1}{128\pi}  \frac{ m_f^2 }{v^2} m_{H_0} \left(1-\frac{2m_f^2}{m_{H_0}^2}\right)\sqrt{1-\frac{4m_f^2}{m_{H_0}^2}} ,
\end{eqnarray}
where the first expression has been simplified to the case $m_{N_1}\sim m_{N_2} \equiv m_N$, and the couplings in the first and second expressions are:
\begin{eqnarray} 
	g_{N}&= & y_r  \,  U_{\nu_R N_1} U_{\nu_R N_2}
	+ y_M\, (  U_{\nu_R N_1} U_{S N_2} + U_{\nu_R N_2} U_{S N_1} )   ,	
\label{yuk_H0-ni-nj} 
\\
\nonumber
	g_{\ell}&= & y_M \, U_{\nu_R\,  N}  \, U_{S \, \nu_\ell} .
\end{eqnarray}
In our natural scenario we assume the Yukawa couplings $y_M$ and $y_r$ to be of order near $10^{-1}$ and the mixings of heavy states $U_{\nu_R N}$ and $U_{S N}$ should be a generic rotation angle, as stated above Eq.~\eqref{rotations}, so we expect $g_N\sim 10^{-1} - 10^{-2}$. 
On the other hand, $g_\ell$ is suppressed by heavy-to light mixing. Consequently, $H_0\to  NN$ should be the largest of the channels into fermion pairs, but even so, it should be below $H_0\to Z' Z'$ according to the above expressions as: 
\begin{equation}
	\label{dec_coc_H0} \frac{\Gamma(H_0\to N_iN_j)}{\Gamma(H_0\to Z'Z')}\sim \frac{|g_N|^2}{\lambda_3}\lesssim 10^{-1} .
\end{equation}

On the other hand, the productions of $H'_0$, $H''_0$ and $H^{\pm}$ through vector boson fusion or bremsstrahlung (Figs.~\ref{Higgs_SM}.b and \ref{Higgs_SM}.c) are extremely suppressed in their couplings  with respect to the SM Higgs by a factor $v_H/v$, and are not produced at all by either gluon fusion or $t\bar t$ fusion (Figs.~\ref{Higgs_SM}.a and \ref{Higgs_SM}.d respectively) as they do not couple directly to quarks. 
The most favourable production of these extra scalars would occur through $Z'$ decay into scalar pairs, provided the channel is kinematically allowed. Even then, only two channels into scalar pairs are not suppressed: 
\begin{eqnarray}
	\nonumber 
\Gamma(Z'\to H^+H^-)&=&\frac{g_1'^2}{3\pi}M_{Z'}\left(1-\frac{4 m_{H^+}^2}{M_{Z'}^2}\right)^{3/2} ,
\\
\label{decay_Z'_SS} 
\Gamma(Z'\to H_0'H_0'')&=&\frac{g_1'^2}{3\pi}M_{Z'}\left(1-\frac{4 m_{H_0'}^2}{M_{Z'}^2}\right)^{3/2}.
\end{eqnarray}
The other modes, namely $Z'\to H_0 H_0$, $Z'\to H_0 H_0'$ and $Z'\to H_0 H_0''$, either vanish identically or are suppressed by the tiny mixing of $Z'$ with the SM $Z^0$ (see Eq.~\eqref{vectormass}).  In our previous analysis of $N$ production through $Z'$ decays leading to Fig.~\ref{fig_BrZp} we did not consider production of scalars, assuming them to be too heavy. Now, to simplify our analysis on scalar production, we will assume the $Z'\to NN$ channels to be absent, with the understanding that in a general case one should consider all scenarios. 

\begin{figure}[h!]
 {\includegraphics[width= 0.5\textwidth]{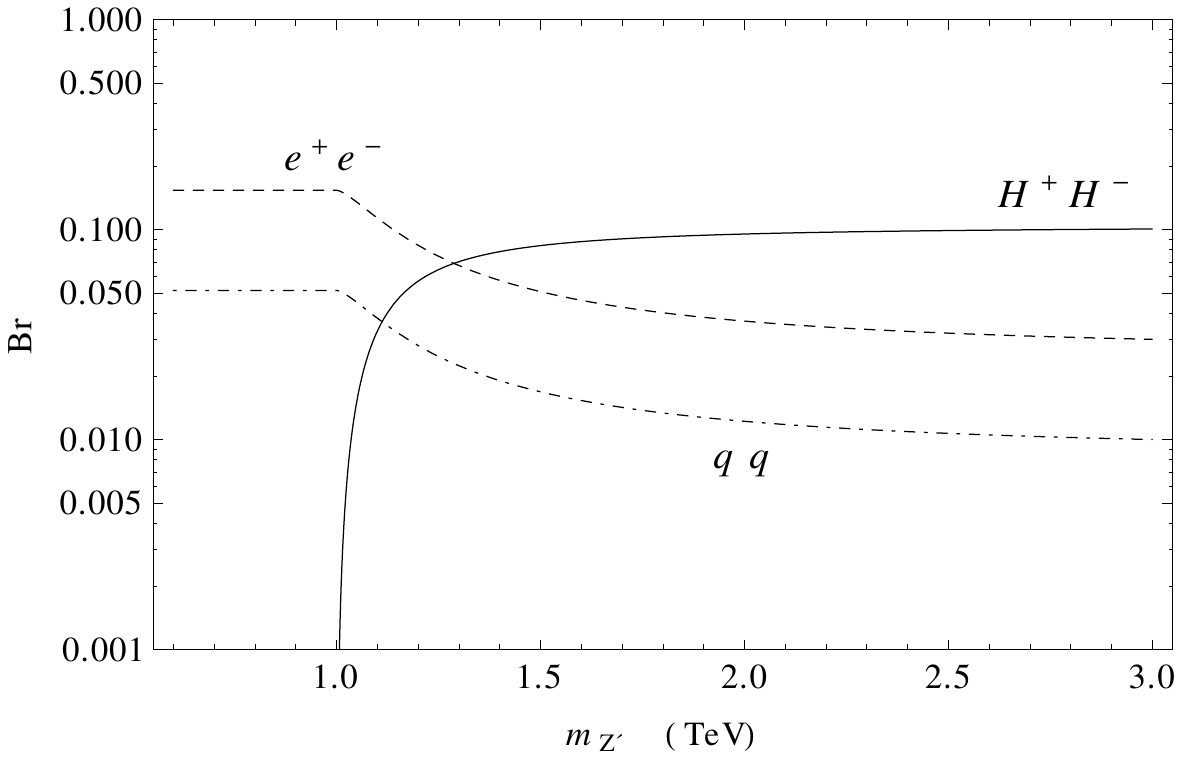}}
	\caption{Branching ratios for $Z'$ decays into extra scalar pairs and SM particles as a function of $M_{Z'}$, for $m_H = 0.5$ TeV.  The $Z'\to NN$ channels are assumed to be absent. The $H_0' H_0''$ production is the same as $H^+H^-$ because  the rates and masses are all the same  in our model. The $\mu^+\mu^-$ and $\tau^+ \tau^-$ channels are the same as $e^+e^-$ and 
	 $q\bar q$ represents any quark-antiquark pair.}
	\label{fig_Br_ZHH}
\end{figure}

Fig.~\ref{fig_Br_ZHH} shows the branching fractions of $Z'$ decays into $H^+H^-$ and $H_0' H_0''$ relative to $Z'$ into SM fermions, for a given value of $m_H$. Scalar pairs clearly dominate the $Z'$ decays as long as they are kinematically allowed. One should keep in mind that the absolute rates, on the other hand, are determined essentially by the $B-L$ gauge coupling  $g_1'$.  In addition, the  observation of these scalars depends on their decay modes. 

Since $H^\pm$, $H_0'$ and $H_0''$ are practically the fields contained in the extra doublet $H$, their main decays are into lepton pairs. Decays into two vector bosons, such as $H^\pm\to W^\pm A$, $W^\pm Z$ or $H_0', H_0'' \to W^+W^-, ZZ$ are highly suppressed by factors of $v_H/v_\chi$. 

Accordingly, for the charged scalars the main decay channels are $H^\pm\to \ell^\pm \nu_\ell$, and
also  $H^\pm\to \ell^\pm N$ if kinematically allowed. The corresponding rates are:
\begin{eqnarray}
\Gamma(H^+\to l^+\nu_\ell )&\simeq & \frac{1}{16\pi}| y_\varepsilon U_{S\nu_\ell}|^2 m_{H^+}\left(1-\frac{m_l^2}{m_{H^+}^2}\right)^3 ,
\\
\nonumber \Gamma(H^+\to l^+N)&\simeq & \frac{1}{16\pi}| y_\varepsilon U_{SN}|^2 m_{H^+}\left[1-\frac{2m_N^2}{m_{H^+}^2}\right]\left[1-\frac{m_N^2}{m_{H^+}^2}\right] .
\label{Hpdecays}
\end{eqnarray}
Here the Yukawa coupling $y_\varepsilon$ is related to the small part $\varepsilon$ of the \emph{linear seesaw} neutrino matrix [see Eq.~\eqref{blocks}]. By our criterium of naturalness,  
$y_\varepsilon$ is not expected to be very small  ($\varepsilon$ should be small because of $v_H$, not because of $y_\varepsilon$).  In turn, $U_{S\nu_\ell}$ and $U_{SN}$ are the elements of the mixing matrix $U_{i\alpha}$, Eq.\eqref{matrixU}, that connect the extra neutrino $S$ with the light and heavy neutrinos, respectively. The former are small, of order $v/v_\chi$, while the latter are generic rotations of order 1.  Therefore, the channels $H^+\to \ell^+N$, if kinematically allowed, will dominate over $\ell^+\nu_\ell$.

On the other hand, the neutral scalars $H'_0$ and $H''_0$ do not even couple to charged leptons, so their only sizeable decay modes are into neutrino pairs, $H_0' , H_0'' \to \nu \nu$ and, if kinematically allowed, into $\nu N$ or $NN$. Consequently these neutral scalars can only be detected if they decay into heavy neutrinos that subsequently decay; otherwise they will be very difficult to detect.

\section{Summary and Conclusions.}

We present a model for neutrino masses based on a linear seesaw scenario where $B-L$ is a local symmetry, the latter included as an extra $U(1)$ in the gauge group. This symmetry is spontaneously broken near the TeV scale. Since we do not want to include effective non-renormalizable operators, then the model requires, besides the Standard Model spectrum,  the inclusion of three neutral fermions per family which are singlets under the SM gauge group, one scalar doublet and one scalar singlet, and the gauge field for $B-L$.  The spontaneous symmetry breaking of this extended gauge group yields masses to light and heavy neutrinos as in a linear seesaw scenario. There also remain a massive $Z'$, and six massive scalars (four neutral and a charged pair).  
The model can naturally accommodate a $Z'$ mass in the range $1- 10\,\TeV$, so its presence could be testable at the LHC, particularly as a dilepton signal \citep{Khalil:2010iu}. 
As any low scale seesaw, the model also contains heavy neutrinos, in this case two per family, with masses also near the TeV scale. The possible signals of these heavy neutrinos, common to all low scale seesaw models, were briefly described. The main distinction from other local $B-L$ seesaw models is the spectrum of four neutral scalars and a pair of charged ones. We required that one of the neutrals must have the mass of the recently discovered 125 GeV state with the phenomenology of the SM, while the rest must remain heavier, 1 TeV or above. This requirement, together with the \emph{linear seesaw} texture of the neutrino mass matrix, imposes some hierarchy conditions on the parameters of the scalar potential, because we seek natural, not finely tuned, scenarios. We thus required that all mass scales in the problem must arise from the pattern of vacuum expectation values of the scalars, and not on tiny Yukawa couplings. We found that this is possible if the parameters of the potential follow some degree of hierarchy, according to: terms that contain purely SM fields, terms with purely extra fields, and terms with both. A subtle point here is the condition that one v.e.v. must be much smaller than the rest, imposed by the linear seesaw texture. This condition can be justified by the fact that the term that sets the small v.e.v. scale explicitly breaks an additional $U(1)$ symmetry in the potential. 
The main phenomenological constraints of the model, which we included in the analysis, are the Higgs mass, current bounds on the light neutrino masses, on heavy bosons masses and on lepton flavour violating processes. Further features of the neutrino sector, such as the mass hierarchy of the light states or the mixing among them, are finer details that do not impose constraints at the level of the construction presented here. Also, it was assumed that the type of $B-L$ violating interactions our scenario could generate in the early Universe are not detrimental to the baryon asymmetry of the Universe.    
Concerning the production of heavy particles at high energy colliders, we found that the best production mode of heavy neutrinos, $N$, is through $Z'\to NN$, provided $M_{Z'}$ is above the $NN$ threshold. Otherwise, $N$ can be produced through $Z'\to \nu_{l} N$, but with a rate at least two orders of magnitude below. Independent of those modes, the existence of this $Z'$ (a $B-L$ remnant) can be detected preferably by its decay into charged lepton pairs, with branching ratios above $10\%$. Concerning the scalar sector, one of them corresponds to the SM Higgs, with mass and phenomenology indistinguishable from the purely Standard Model. The other scalars, three neutral and two charged, have no direct coupling to quarks and very suppressed to SM leptons. Thus their main production at hadron colliders is through $Z' Z'$ fusion, either real or virtual, or by $Z'$ decay into scalar pairs, provided the corresponding channel is open.  The charged scalars, if produced, will decay leptonic, $H^+\to \ell^+ N$ and $\ell^+ \nu_\ell$, preferably the first mode if open, while the second is suppressed by heavy-to light neutrino mixing. Of the three extra neutral scalars, the one composed mostly of the singlet field, which we call $H_0$, is the most visible of them: it will decay mainly into $Z' Z'$ and $N N$ (if allowed), with all other modes suppressed by at least one order of magnitude.    The other two neutral scalars, $H'_0$ and $H''_0$, are much more difficult to detect, as their production is mainly through $Z'\to H'_0 H''_0$ and their dominating decays are into neutral lepton pairs, $NN$, $\nu N$ and $\nu \nu$. 

Our final and main conclusion is that the local $B-L$ versions of the linear and inverse seesaw have similar phenomenology in the heavy neutrino and $Z^\prime$ sectors, and therefore it will be difficult to distinguish experimentally between the two models from these sectors. The main difference between these two models is actually in the scalar sector, in particular in the appearance of charged heavy scalars in the linear seesaw, which are absent in the inverse seesaw. Conversely, if the heavy scalars are too heavy to be directly produced at the LHC, the distinction between inverse vs. linear seesaw will remain a challenging task for experiments.

\acknowledgments
We are grateful to Zackaria Chacko, Alexander S.~Belyaev and Julian Heeck for valuable  comments. This work was supported in part by Fondecyt (Chile) grant 1130617.

\bibliography{biblio.bib}	
\bibliographystyle{apsrev4-1}
%
\end{document}